\documentclass[12pt,preprint]{aastex}

\begin{document}


\title{Circumnuclear stellar population, morphology
and environment of Seyfert 2 galaxies: an evolutionary scenario}

\author{Thaisa Storchi-Bergmann}
\affil{Instituto de F\'\i sica, UFRGS, Campus do Vale, CP\,15051, Porto
Alegre, Brasil}
\email{thaisa@if.ufrgs.br}

\author{Rosa M. Gonz\'alez Delgado}
\affil{Instituto de Astrof\'\i sica de Andaluc\'\i a (CSIC), Apdo. 3004,
18080, Granada, Spain}
\email{rosa@iaa.es}

\author{Henrique R. Schmitt\altaffilmark{1}}
\affil{National Radio Astronomy Observatories, P.O. Box 0, Socorro, NM87801}
\email{hschmitt@aoc.nrao.edu}

\author{R. Cid Fernandes\altaffilmark{2}}
\affil{Department of Physics \& Astronomy, Johns Hopkins University,
Baltimore, MD, 21218}
\email{cid@pha.jhu.edu}

\author{Timothy Heckman}
\affil{Department of Physics \& Astronomy, Johns Hopkins University,
Baltimore, MD, 21218}
\email{heckman@pha.jhu.edu}

\altaffiltext{1}{Jansky Fellow}
\altaffiltext{2}{Gemini Fellow}

\begin{abstract}

We investigate the relation between the characteristics of the 
circumnuclear stellar population and both the 
galaxy morphology and the presence of close 
companions for a sample of 35 Seyfert 2 nuclei. 
Fifteen galaxies present unambiguous signatures 
of recent episodes of star formation within $\approx$300 pc 
from the nucleus. When we relate this property with the Hubble 
type of the host galaxy, we find that the incidence of recent 
circumnuclear star formation increases along the Hubble 
sequence, and seems to be larger than in non-Seyfert galaxies for 
the early Hubble types S0 and Sa, but similar to that in non-Seyfert
galaxies for later Hubble types. Both in early-type and late-type
Seyferts, the presence of recent star-formation
is related to the galaxy morphology in the inner few kiloparsecs,
as observed in HST images through the filter F606W by Malkan et al., 
who has assigned a late ``inner Hubble type'' to most Seyfert 2s 
with recent nuclear star-formation. This new classification
is due to the presence of dust lanes and spiral structures in the inner region.
The presence of recent  star formation in Seyfert 2 nuclei is also related to interactions: among the 13 galaxies of the sample with close companions or in mergers, 9 have recent star formation in the nucleus. These correlations between
the presence of companions, inner morphology and the incidence of recent
star formation suggest an evolutionary scenario in which the interaction
is responsible for sending gas inwards which both feeds the AGN and triggers 
star-formation. The starburst then fades with time and the composite
Seyfert 2 + Starburst nucleus evolves to a ``pure'' Seyfert 2 
nucleus with an old stellar population. This scenario can reconcile
the hypothesis that the active nucleus in Seyfert galaxies 
is triggered by interactions with the results of previous studies which find
only a small excess of interacting galaxies in Seyfert samples
when compared with non-Seyfert samples. The large excess can only
be found early after the interaction, in the 
phase in which a composite (Seyfert + Starburst) nucleus
is observed.

 
\end{abstract}

\section{Introduction}

The recent evidence for a proportionality between galactic bulges and nuclear black-hole masses (Magorrian et al. 1998, Ferrarese \& Merrit 2000;
Gebhart et al. 2000), and the fact that supermassive black-holes seem to
be present in the nuclei of most present-day galaxies (Ho 1999),
point to a ``starburst-AGN connection'' operating at the 
epoch of galaxy formation. Circumstantial evidence for this connection
is, for example, the quasar
host images obtained with the Hubble Space Telescope (Bahcall et al. 1997),
showing distorted morphologies due to interactions, characteristic
of luminous starbursts in the near Universe, and the Ultra-luminous Infrared
Galaxies (ULIRG), with quasar luminosities (L$_{IR}>10^{12}$L$_\odot$).
The latter are in most cases star-forming merger systems argued to be
the  initial dust-enshrouded stage of a quasar.  Spectral signatures of
an ageing starburst have indeed been found in a few QSO's  and
ULIRGS (Brotherton et al. 1999;  Canalizo \& Stockton 2000), interpreted
as being objects in the transition  phase between the starburst and
the QSO. 

In the near Universe, starburst galaxies and
active nuclei share a fundamental characteristic: both are dependent on
gas to fuel the birth of new stars in the first case and 
to feed the nuclear black-hole in the second.
If there is a gas flow to the center, it may trigger
star-formation. This is the essence of the hybrid models proposed
by Perry \& Dyson (1985) and Norman \& Scoville (1988). More recently,
Collin \& Zahn (1999) argue that  star-formation events can occur 
as far inwards as in the outskirts of the accretion
disk, where the gas is gravitationally unstable.

Observational evidence for nuclear starbursts around nearby AGN includes
the works of Terlevich, Di\'az \& Terlevich (1990), Heckman et al.
(1997) and Gonz\'alez Delgado et al. (1998). In a recent work, Aretxaga et al.
(2001) reported the detection of prominent Balmer absorption lines in 
six radio-galaxies, attributed to stars younger than 1\,Gyr.
Cid Fernandes \& Terlevich (1992,
1995) have shown how the presence of a nuclear starburst could
solve the ``FC2 problem'' (Tran 1995a,b,c), the unpolarized blue light present
in the spectra of many Seyfert 2 galaxies. Cid Fernandes, Storchi-Bergmann \& 
Schmitt (1998), Storchi-Bergmann, Cid
Fernandes \& Schmitt (1998) and Schmitt, Storchi-Bergmann \& Cid
Fernandes (1999) showed that the FC2 problem in the optical
can be solved if one takes into account the fact 
that the  nuclear stellar population
of Seyfert galaxies is varied, and cannot in most cases be
represented by an elliptical galaxy template. Intermediate age 
($\approx$10$^8$yr) and young stars are responsible 
for the excess optical light in many cases. 

Are nuclear starbursts ubiquitous in Seyfert 2 galaxies? 
In order to answer this question it is necessary
to quantify the frequency of recent star formation episodes in
or around Seyfert 2 nuclei. This has been done 
in 2 recent works, by Gonz\'alez Delgado, Heckman \& Leitherer 2001
(hereafter GD01) and Storchi-Bergmann et al. 2000 (hereafter SB00), 
who have analyzed the near-UV spectra of 20 Seyfert 2 galaxies each.
Through spectral synthesis techniques, GD01 and SB00 have
quantified the contribution of old, intermediate and young
stellar components to the spectra.  Unambiguous signatures 
of recent star formation have been found in 50\% of the 
galaxies of the sample of GD01,
while SB00 have found such signatures in 30\%  of their sample.
In another 30\% of the galaxies of the two samples,
a power-law component, contributing less than 30\% of the flux
at $\lambda$4020 was necessary to reproduce the near-UV continuum.
SB00 has  shown that this component cannot
be distinguished from the continuum produced by a starburst
of 10\,Myr or younger, for such small flux contributions.
SB00 has called this component PL/YS.
If this latter component were entirely due to young stars,
then the fraction of Seyfert 2 with recent star formation
would increase to 80\% for the northern sample and to 60\%
for the southern sample.

In order to investigate if the incidence of recent 
star-formation events is larger in the nuclei of
Seyfert 2 galaxies than in non-Seyferts
(implying a starburst-AGN connection),
it is necessary to compare the results obtained for the Seyferts with those
for non-Seyfert galaxies. This is the goal of this work.
In  Section 2 we describe the sample, 
in Section 3 we discuss the relation between
the stellar population characteristics and the Hubble type of the
galaxy, in Section 4 we compare the results for the Seyferts with
those of non-Seyfert galaxies with the same Hubble type, in section 5
we discuss the role of interactions, in section 6 we discuss the 
relation between the stellar population and the inner morphology 
of the galaxy and in Section 7 we present the conclusions of this work.

\section{Sample}

We use for this work the combined samples of SB00 (hereafter the southern sample) and GD01 (hereafter the northern sample). Our goal is to collect
a sample of Seyfert 2 galaxies in the local Universe,
spanning a range of morphological and
environmental characteristics, whose spectra have been
observed with similar instrumentation
and for which the stellar population has been studied using similar techniques,
such that a common characterization of the stellar population can be used.

The southern sample comprises approximately 40\% of the Seyfert 2 galaxies from the catalogue of V\'eron-Cetty \& V\'eron (2000) 
with redshift  $z<0.02$ and luminosity in the [OIII]$\lambda$5007 
emission-line $L_{[OIII]}>10^{40}$ergs~cm$^{-2}$ s$^{-1}$, 
which could be observed from the southern hemisphere.
In addition, it contains the galaxies CGCG 420-015, MCG-5-27-13 which
obey the [OIII] luminosity criterium but are somewhat more distant,
with z=0.029 and z=0.024, respectively. The southern sample 
can be considered a local sample, selected on the basis of the central source luminosity (via $L_{[OIII]}$).

The northern sample was selected according to the flux of the central 
source: it comprises approximately 80\% of the Seyfert 2 galaxies from the Whittle (1992) sample, observable from the northern hemisphere and which have 
fluxes $F_{[OIII]}>0.6\times 10^{-12}$ergs~cm$^{-2}$s$^{-1}$ and/or 
$F_{1.4 GHz}>80$ mJy.
Besides 5 galaxies in common with the southern sample, the northern sample comprises other 7 galaxies which also obey the selection criteria of the southern sample, but in addition contains 8 galaxies with $z>0.02$,
which have on average larger central source luminosities.

In summary, the combined sample  contains 25 of the closests Seyfert 2
galaxies ($z<0.02$) plus other 10 with $0.02<z<0.05$. 
The common selection criterium for all the galaxies is the 
luminosity of the central source, which is higher than a lower limit which produces $L_{[OIII]}>10^{40}$ergs~cm$^{-2}$s${-1}$. 
We will regard this sample as representative of 
nearby Seyfert 2 galaxies. As it was
not selected by any property related to the stellar population, galaxy morphology or environment, it is suitable to explore the relation 
among the latter three properties. 

The galaxies' properties relevant for this work are listed in Table 1,
including radial velocities, absolute magnitudes (for H$_0$=75~km~s$^{-1}$
Mpc$^{-1}$, used throughout this paper), scale at each galaxy, Hubble type as listed in RC3 (de Vaucouleurs et al. 1991) or in NED 
(NASA/IPAC Extragalactic Database) and an
``inner'' Hubble type proposed  by Malkan, Gorjian \& Tam (1998), 
on the basis of HST images (see discussion in Sec. 5).
In the last column, we list a number representing a characterization of
the stellar population at the nucleus, based on the analysis of GD01 and SB00,
as described below.  

We have divided the stellar population in three categories,
represented by: the number `1'  when the nuclear spectrum presents 
unambiguous signatures of recent star formation (younger than 500\,Myr); 
these cases are also called composites (Starburst+Seyfert), as their emission-line spectra have line ratios intermediate between
those of Starbursts and Seyferts (e.g. Cid Fernandes et al. 2001);
the number `2' when the stellar population is dominated by components
older than 1\,Gyr; the number `3' when  a PL/YS continuum 
is necessary to reproduce the spectra in the near-UV. 
The sample comprises 15 composites ($\approx$40\% of the sample), 
10 dominated by an old stellar population ($\approx$30\%) and
10 ($\approx$30\%) for which there is a need for a blue continuum which
can be both due to a very young stellar population (but for which
it is not possible to detect stellar absorption features) or
to a featureless continuum of non-stellar origin.

\section{Hubble types} 

\begin{figure}
\plotone{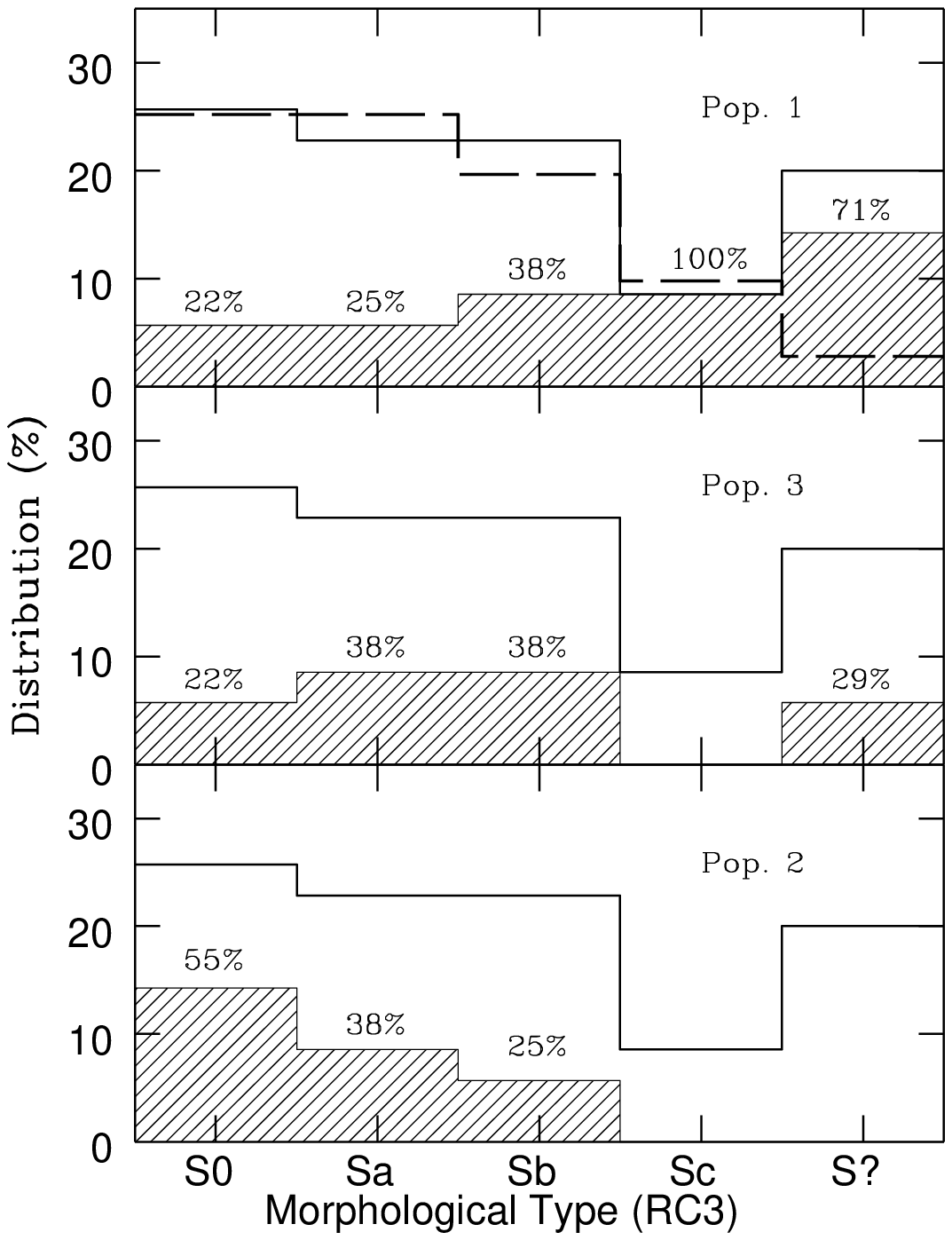}
\caption{Open histograms: percent distribution of Hubble types for the present 
Seyfert 2 sample (continuum line) compared with that of Schmitt et al. 
(2001) (dashed line in the top panel). Hatched histograms:
the fractions of Seyfert 2 galaxies belonging to each stellar population
category. From top to bottom, categories 1 (composites),
3 (PL/YS - blue continuum with uncertain origin) 
and 2 (old stellar population). The fractions are
also labeled with corresponding percentages within each Hubble type.}
\label{fig.1}
\end{figure}

We show in Fig.\ref{fig.1}, histograms of the Hubble types of our Seyfert 2 sample,
where we have grouped the S0/a galaxies with the Sa, the Sab with 
the Sb and the Sbc with the Sc. A few galaxies have uncertain
Hubble types, due to both a distorted morphology, or
to the fact that the galaxies are too distant to allow a 
morphological classification based on available images. We have 
grouped the latter galaxies in Fig. 1 in a 
column beyond that of Sc, identified
as S?. The open histogram corresponds
to the whole sample, and the hatched histograms to 
subsamples separated according to the stellar population categories described
in the previous section: from top to bottom, categories `1', `3' and `2'.  

From Fig.\ref{fig.1} it can be concluded that the dominant Hubble types are 
S0 and Sa, closely followed by the Sb, then the number of galaxies
drops by more than 50\% for the Sc, and increases again for the 
uncertain types. In order to evaluate if the morphological type
distribution of our sample is representative of a better defined Seyfert
sample, we compare it with that of
Schmitt et al. (2001, hereafter S01). Their sample is selected on the basis
of the 60$\mu$m infrared luminosity of the galaxies, a property believed
to be isotropic, and contains 
approximately twice as many Seyfert 2 galaxies as our sample, with 
18 S0, 18 Sa, 14 Sb, 7 Sc and 2 of uncertain type.
Their distribution of Hubble types is also shown in Fig. 1
as a dashed histogram. It is very similar to that of our sample,
although has a much smaller number of uncertain types. 
We attribute this difference to the fact that 5 of the 7 S? galaxies in
our sample are the most distant ones, with much larger distances
than those of S01 sample, being difficult to classify.
Excluding these galaxies, the number of galaxies with uncertain
classification in our sample is not significantly different from
that in S01 sample. Due to this difference, in Fig. 1 we have normalised the
S01 distribution to ours excluding the uncertain types.

Regarding the stellar population category, 
the composites (category `1') comprise approximately
20-25\% of the S0 and Sa, $\approx$40\% of the Sb, 100\%
(although there are only three) of the Sc and $\approx$70\% of the S?. 
The galaxies dominated by the old
stellar population (category `2') are clearly concentrated towards the
early types, with increasing percentages 
from $\approx$25\% for the Sb to $\approx$60\%
for the S0. The galaxies with the PL/YS component (category `3') 
are approximately evenly distributed among the different Hubble types, 
comprising approximately 30\% of the galaxies. 

Now let's consider the possibility that the PL/YS continuum  
is produced by young stars. As   
discussed in Cid Fernandes et al. (2001), this could happen due
to a contrast effect: the nuclear starburst would be too
faint as compared with the contribution of the bulge, and 
the photospheric lines of the young stars would not be detected.
The presence of a faint starburst in these cases would
only be detected  via the blue continuum and/or the dilution
produced in the near-UV absorption features of the bulge stellar population,
as observed in the galaxies of category 3. In order to take into account
this possibility, we present in Fig. 2 a revised 
version of the top panel of Fig. 1, where
we assume that all nuclei with stellar population of category 3 
are also composites.

\section{Comparison with non-Seyfert samples}

A similar stellar population study to those of SB00 and GD01
was performed for local galaxies by
Bica \& Alloin (1987) and Bica (1988, hereafter B88).
We have used B88 sample as a comparison sample for the distribution of the stellar population characteristics among the different Hubble types. The comparison is relevant because the stellar population characterization is essentially the same as that we have for our sample and
B88 sample is dominated by non-Seyfert nearby galaxies ($z<0.02$).
His goal in assembling the sample was to have a representative 
number of galaxies spanning the Hubble types E-Sc and absolute magnitudes 
$-22<M_B<-16$. In addition, 8 Seyferts and 2 Starburst galaxies were also included. In order to compare our results with those of B88, 
we have excluded from his sample the latter and the elliptical galaxies.
The remaining subsample comprises 117 spiral galaxies, 
distributed as 32 S0's, 25 Sa's, 26 Sb's and 34 Sc's.
B88 has avoided including in his study galaxies with uncertain morphology,
so we cannot complete the S? column with his results. The percent histogram of
Hubble types of the B88 sample is shown in the top panel of Fig.\ref{fig.3}. 
On the basis of the stellar population analysis of B88, we were able to
classify the stellar population of his work within the categories `1' and `2' 
described in Sec. 2. The corresponding distributions of stellar population
categories are shown as hatched histograms in the first and third panel
(from top to bottom) of Fig.\ref{fig.3}.

In order to check if the distribution of morphological types in the 
B88 sample is representative of those of a ``complete sample'' of
non-Seyfert galaxies, we have compared it with that of  
the Palomar spectroscopic survey 
of nearby galaxies by Ho, Filippenko \& Sargent (1997, hereafter 
HFS97). This survey provides a representative sample of the galaxies in the near-Universe, and can thus be used as a reference for the
distribution of galaxies among the different Hubble types.
The HFS97 sample comprises most of the northern galaxies brighter 
than B$_T$=12.5 mag, with a total number of 486 galaxies, among which
57 are ellipticals and 52 are Seyferts. 
The percent distribution of Hubble types of the HFS97
survey is shown in the second panel (from top to bottom)
of Fig.\ref{fig.3}, after excluding the elliptical and Seyfert galaxies.
As HFS97 include also galaxies with morphological 
types beyond Sc, we construct the S? bin
adding all galaxies with these later classifications. But in order to make
the HFS97 distribution comparable to that of B88, we have normalized
to the total number of galaxies from S0 to Sc, excluding the S?. 
HFS97 do not perform a stellar population study as B88, but
have made a careful analysis of the emission-line spectra, classifying
the galaxies as LINERs, Seyferts and HII nuclei. Considering that an
HII region spectrum is a tracer of very recent star-formation,
we tentatively use this classification as indicative of
a stellar population of category 1. The hatched histogram in
the second panel (from top to bottom) of Fig.\ref{fig.3} 
shows the distribution of the HII region
nuclei within the Hubble types of the HFS97 sample.

Before comparing the Seyfert sample with the two above non-Seyfert samples,
we have also checked if the three samples span similar ranges in
absolute magnitude M$_{B_T^0}$. Fig.\ref{mag} shows 
that this is indeed the case for
the bulk of the galaxies in each sample.
The  distribution in absolute magnitude of the Seyferts 
is more similar to that of B88 than to that of HFS97, considering
only the galaxies more luminous than $M_{B_T^0}=-19$, as that of HFS97
is sistematically shifted relative to the distribution of Seyferts
to less luminous galaxies by $\approx$ 
0.5\,mag. Below (lower luminosity) $M_{B_T^0}=-19$, 
there are low luminosity tails in HFS97 and B88 distributions, 
not present in the Seyferts distribution. However, these 
low luminosity galaxies comprise only 12\% of B88 and 17\% of HFS97 samples.
We thus conclude that the Seyfert sample  can be considered 
comparable to that of B88 in terms of absolute blue magnitude distribution,
but both are shifted towards higher luminosity (by 0.5\,mag) 
in comparison to the complete sample of HFS97.

From Fig.\ref{fig.3} it can be observed that both B88 and the HFS97 surveys 
present an approximately uniform distribution
of Hubble types along the sequence S0 to Sc, 
and that the number of galaxies with 
Hubble type later than Sc in the Palomar survey is approximately 
half the number of Sc galaxies. 
 
A comparison between Fig.\ref{fig.3} and Fig.\ref{fig.1} shows that the main difference between the Hubble type distributions of our 
Seyfert sample and those of the non-Seyfert
samples is the smaller proportion of Sc's among the Seyferts 
(approximately half that of the non-Seyferts).
Another difference is the relative number of galaxies with uncertain or 
peculiar morphology, which is larger in the Seyfert sample 
when compared with that of the Palomar Survey. 

Regarding the stellar population categories, the incidence of recent star-formation in normal galaxies increases from less than 10\% for S0
to 70-80\% for Sc and S?. The incidence of recent star formation
seems to be somewhat larger in the HFS97 sample than in the B88 sample.
We attribute this difference to the distinct methods used
to trace the young stellar population in the two samples, 
which favors the detection of fainter bursts of star formation when
tracing them by the emission lines (HFS97), considering also that
the detection limit of emission lines of HFS97 ($\approx$0.25\AA)
is lower than that of Bica's observations ($\approx$2\AA; 
Bonatto, Bica \& Alloin 1989).
For the Seyferts, we identify a similar trend of increasing 
incidence of recent star formation towards later 
Hubble types. When compared with the B88 sample, the early-type Seyferts
S0 and Sa present a larger incidence of recent 
star formation than the non-Seyfert galaxies, but,
in the case of Sa, the incidence of star-formation is similar
to the percentage of HII nuclei in the HFS97 sample.
Although the number of galaxies of the present sample is 
still small for a firm conclusion on this issue, a recent work by Raimann et
al. (2001) has revealed two additional cases of nearby Seyfert 2 nuclei 
with recent star formation in S0 galaxies, supporting the conclusion
that, at least for the S0 hosts, the Seyfert 2 nuclei show a larger
incidence of star formation than those of the non-Seyfert galaxies.

\begin{figure}
\plotone{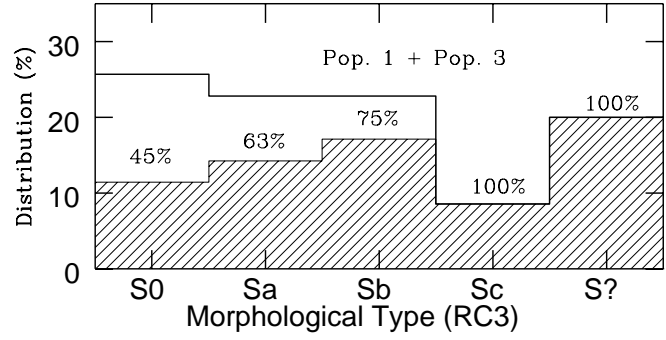}
\caption{The open histogram shows the distribution of Hubble types
as in Fig. 1, while the hatched histogram shows the fraction
of galaxies for each Hubble type having a stellar population
of category 1 or 3.}\label{fig.2}
\end{figure}

\begin{figure}
\plotone{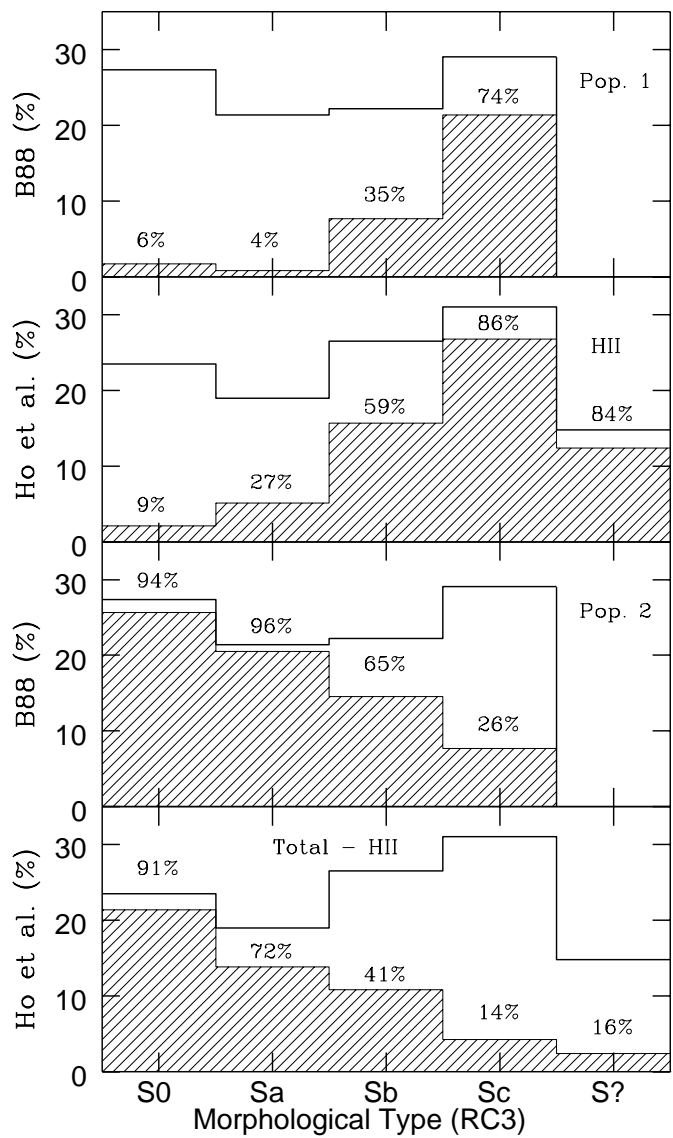}
\caption{Open histograms show the percent distribution of Hubble types 
for two control samples of normal galaxies: the sample of Bica 
(B88) and that of Ho et al. (1997). 
Hatched histograms show the fraction of normal galaxies
with recent star formation in the two upper panels, 
and with old stellar population 
in the two bottom panels. These fractions are also labeled with
corresponding percentages within each Hubble type.}\label{fig.3}
\end{figure}

\begin{figure}
\plotone{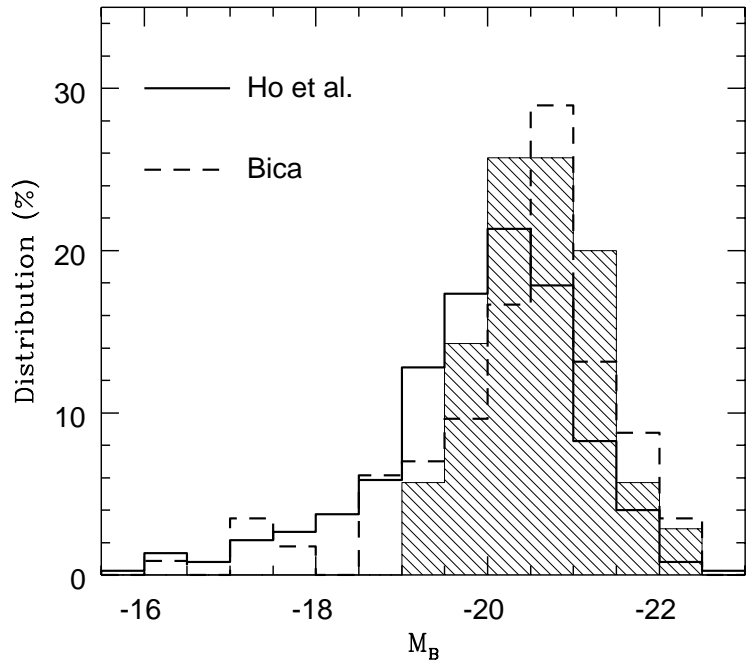}
\caption{The hatched histogram shows the distribution of 
absolute blue magnitudes of the Seyfert sample, which can
be compared with that of B88 (dashed) and HFS97 (heavy
continuous line) samples.}\label{mag}
\end{figure}

We can also compare Fig.\ref{fig.3} with Fig.\ref{fig.2}, in which we have assumed that the ambiguous blue continuum of 
category 3 is also due to young stars.
If this is the case, it is clear that the incidence of
recent star-formation in Seyfert 2s would be larger 
than in non-Seyfert galaxies
for all Hubble types. This conclusion would still hold
even if only half of the population of category 3 is due to young
stars. It is thus very important to investigate further the nature of the PL/YS
continuum. Cid Fernandes et al. (2001), for example,
has shown that, for the same sample studied here, the galaxies show a 
continuum of infrared (IRAS) luminosities L$_{IR}$, the more luminous being 
the composites (Seyfert 2 + Starburst). 
They have also found that L$_{IR}$ is correlated
with the fraction of the optical continuum due
to blue light, the galaxies from category 2 being the less luminous
in the IR and showing almost no blue light, followed by the
PL/YS cases up to the composites, which show the larger
fractions of blue light in the optical continuum and larger L$_{IR}$s.  
Cid Fernandes et al. results thus suggest that the same phenomenon is
occurring in the whole sample with a range of luminosities, and that
the PL/YS could be indeed be due to young stars, not detected because of
the contrast effect discussed in the previous section.

The fraction of galaxies dominated by old stellar population increases towards 
the early Hubble types in both Seyferts and non-Seyferts, although this fraction
is smaller in the early-type Seyferts when compared to the non-Seyfert
galaxies. This result is due both to the larger incidence of 
composites in S0 and Sa Seyferts discussed above, 
as well as to the $\approx$30\% incidence of galaxies with PL/YS continuum.

\section{Galaxy structure in the inner few kiloparsecs}

Malkan, Gorjian and Tam (1998, hereafter MGT), have recently published the results of an imaging survey of Seyfert and Starburst galaxies
using the WFPC2 aboard HST and the broad-band
filter F606W (which covers the spectral region $\approx$4700-7200\AA). 
Based on these images, which show a wealth of 
fine structure not detected in previous ground based images,
MGT have assigned a Hubble type for the inner region of each galaxy. 
This ``Inner Hubble Type'' is listed in column 6 of Table 1.
A dash identifies the galaxies not observed in the imaging survey of MGT.
Since these classifications are based on images 
obtained with the Planetary Camera,
which has a field-of-view of 37$^{\prime\prime}$, they refer only to the
inner few kiloparsecs of the galaxies, except in the most distant
cases. 

\begin{figure}
\plotone{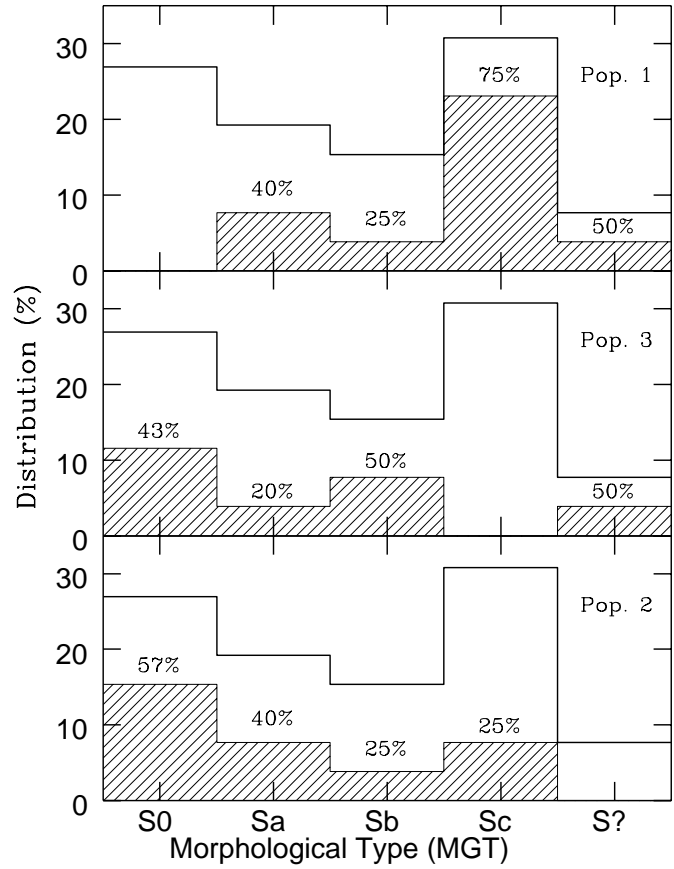}
\caption{Open histograms show the distribution of MGT types (the ``inner Hubble 
types'') of the Seyfert 2 sample, while the hatched histograms show the fraction
of galaxies belonging to each stellar population category. From top to bottom:
categories 1, 3 and 2. The fractions are also labeled with corresponding
percentages within each Hubble type.}\label{fig.4}
\end{figure}

We show in Fig.\ref{fig.4} an histogram with the distribution of our sample 
according to the Inner Hubble types of MGT (hereafter MGT type). 
This figure shows that many galaxies from our sample, in spite of having 
an early RC3 Hubble type, have a late MGT type. Interestingly, most cases 
of composites (category 1)  have a late MGT type, indicating a relation between the MGT type of the Seyfert and its stellar population.
A comparison between Fig.\ref{fig.4} and Fig.\ref{fig.3} shows that
the frequency of recent star-formation in Seyfert 2's of Sc MGT type is
remarkably similar to that in non-Seyfert galaxies of Hubble type Sc.
The ``Sc morphology'' thus seems to be an indicator of the presence
of recent star formation. 

The dominant old population (category 2) is concentrated towards early MGT
types, in this case in agreement with the relation with the large scale Hubble
type. The stellar population category 3 shows a similar 
behavior to that of category 2.

The question we should now answer is: 
what physical characteristics are behind the ``Sc 
morphology'' in F606W images? Such images are tracers of both
the stellar distribution and dust. The dust distribution around
Seyfert 2 nuclei has been recently studied by Martini \&
Pogge (1999) combining F606W images with near-infrared 
NICMOS H images to construct V-H color maps. These authors have
shown that the nuclear regions of Seyfert 2 galaxies are very rich in gas
and dust, and have found nuclear
spiral dust lanes on scales of a few hundred parsecs in 20 galaxies
out of a sample of 24 Seyfert 2s. They suggest that 
these spiral dust lanes are the channels by which gas from the
galaxy disks is being fed into the central engines.
By comparing the F606W images with the
V-H maps of Martini and Pogge, it can be concluded that the structure seen
in the F606W images is mainly due to dust. In addition, 
as the F606W filter includes emission lines such as
H$\alpha$ -- particularly strong in star-forming regions --
large amounts of gas emitting H$\alpha$ may
also contribute to the morphology of F606W images.

Based on the above, one possible interpretation for the late type 
morphology of the inner regions of the Seyfert 2
galaxies with recent star formation is that these galaxies
are particularly rich in emitting gas and associated dust,
such that star formation can be triggered.
A large amount of gas and dust is normally observed
in galaxies with Hubble type Sc, 
which are known to be gas rich and to have a large 
incidence of nuclear starburts. Thus it is not a surprise that
the three galaxies of our sample with Hubble type Sc show recent star 
formation at the nucleus. 
What is new here is the case of the Seyfert 2 nuclei
with earlier Hubble type hosts and recent star-formation. These galaxies
also seem to be particularly rich in gas in the inner few
hundred parsecs around the nucleus to allow the triggering
of star-formation. This interpretation is 
supported by the results of the recent work
of Cid Fernandes et al. (2001) on the same sample studied here,
who have shown that the Seyfert 2 nuclei with clear signatures of recent 
star-formation are the ones with larger far-infrared luminosities,
consistent with stronger emission by starlight-heated 
dust in these galaxies.

\section{The role of interactions}

Having concluded that the Seyfert 2 nuclei
with recent star-formation are particularly rich in gas and dust
in the inner regions, how is the gas transported to the
central region of these galaxies? The problem of how the gas
looses its angular momentum in this process has been discussed by
many authors and several mechanisms have been suggested.
One of these mechanisms is the interaction
between galaxies (Gunn 1979; Hernquist 1989; Hernquist \& Mihos 1995),
which can also trigger star formation before the nucleus is fed 
(Byrd et al. 1986; Byrd, Sundelius \& Valtonen 1987; Lin, Pringle \& Rees
1988). Large scale bars can also remove angular momentum 
via gravitational torques, making the gas to fall inwards
(Shlosman 1992). The latter mechanism seems to be particularly
relevant in triggering nuclear starbursts, which  
preferentially occur in barred hosts (Heckman 1980, Balzano 1983,
Kennicut 1994).
 
Structural distortions which could have been produced by interactions and bars 
have been indeed found by the pioneer work of  Simkin, Su \& Schwarz (1980).
Dahari (1984) found an excess of companions in Seyferts when compared
with a control sample of field galaxes, although the excess is small.
This result has been recently
confirmed by Rafanelli et al. (1995). 
Regarding bars, recent studies using near-IR images
(e.g. Mulchaey \& Regan 1997) did not find more bars 
in Seyfert's when compared with a control sample; the main structure they
found is a nuclear spiral, in agreement with the work
of Martini \& Pogge (1999) discussed in the previous section,
and with the MGT classification. On the other hand,
Knapen, Schlossman \& Peletier (2000), making a careful 
match between the properties of a Seyfert sample with that of
a comparison sample, do find a small excess of
bars in Seyfert's. In our sample, we do not find a relation between the
stellar population properties and the presence of a bar
(RC3 classification SB). A more meaningful study of
such relation can only be done through analysis of near-IR 
images of all galaxies of our sample, not presently available.



We have checked the relation between interactions and
the stellar population category in our sample searching
for companions around our galaxies using NED and the
Digitized Sky Survey (DSS) plates, and found obvious companions and/or
signatures of interactions/mergers in 13 cases.
The criteria used to identify companions include: the proximity
in the sky -- companion should be closer that a few galaxy diameters
of the Seyfert; the difference in radial velocity should be
smaller than $\approx$300 km s${-1}$; the difference in magnitude
should be smaller than 3 mag; the presence of 
apparent tidal distortions in the galaxies images. 
We list in Table 2 the 13 interacting galaxies together with the available
information on the companion galaxies: name, distance from the Seyfert 
in kiloparsecs, radial velocity and difference in absolute
magnitude (magnitude of the companion
minus that of the Seyfert). There are two cases of mergers, in
which it is not possible to separate the companion, and four
cases in which the Seyfert belongs to a group. In the latter cases,
we have included in Table 2 the data for the galaxy from the
group which is closest to the Seyfert. 

There is only one case
for which it was not possible to identify the companion:
NGC\,7130. In the DSS, there seems to be two small companions: the one
at 15.7\,kpc\,NW listed in Table 2, plus another at 9.4\,kpc\,SW
(Gonz\'alez Delgado et al. 1998).
Shields \& Filippenko (1990) have obtained better quality images
and report tidal distortion at faint light levels which could
have been the result of interaction with the small galaxy at
NW, which they also observe in their images. The other
possible galaxy to the SW is not observed. A spectrum of
this small galaxy to NW would be necessary in order to 
conclude if it is indeed gravitationally bound to NGC\,7130.

For reference and comparison, we also include
in Table 2 the radial velocity and absolute magnitude of the corresponding Seyfert galaxy, as well its stellar population category.
All the above cases of interactions have been previously 
reported in the literature, and we thus list the corresponding references
in the last column of Table 2. It can be observed from the Table that,
for most cases the available data -- such as proximity to the Seyfert
and similar radial velocity supports
a physical association between the Seyfert and the companion.
From Table 2, it can also be concluded that the companions are
usually less luminous than the
Seyferts, consistent with the theory by Hernquist \& Mihos (1995), in 
which  minor mergers induce radial inflows which accumulate large quantitites
of interstellar gas in the nuclear regions of the host disks,
which can then feed the nuclear blackhole.

\begin{figure}
\plotone{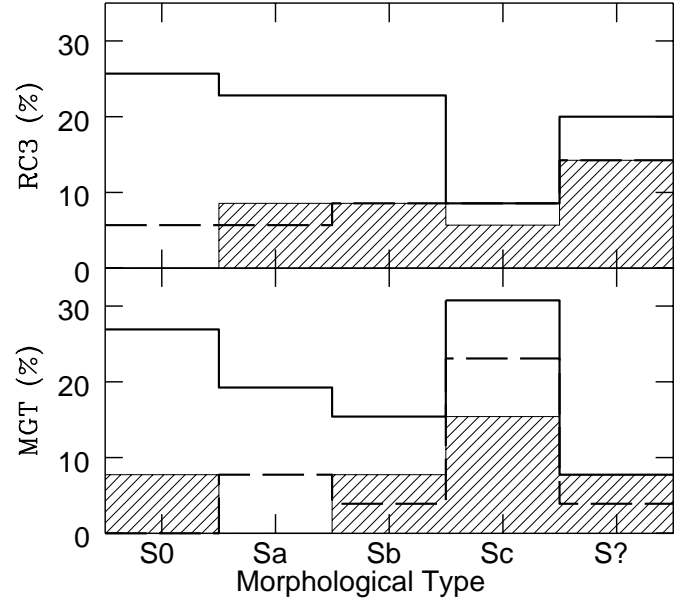}
\caption{The histograms of Hubble types (upper panel) 
and of the inner MGT types (bottom panel) for the Seyfert 2 
sample together with the fraction of galaxies
with close companions (hatched histograms) and 
with recent star formation (dashed histograms).}\label{fig.5}
\end{figure}

The proportion of 13 interacting galaxies out of a sample
of 35 Seyfert 2 galaxies 
is similar to that of the larger Seyfert sample
of Schmitt et al. (2001), selected 
on the basis of the 60$\mu$m luminosity.
Segregating the different stellar population categories,
among the 15 nuclei with recent star formation (category 1,
or composites), 
9 have close companions. In comparison, of the 20 nuclei that do not show
recent star formation, only 4 have close companions. 
Using another perspective, inspection of Table 2 shows that, 
of the 13 galaxies with close companions or in groups,
9 have recent nuclear star-formation. This indicates a relation 
between the starburst activity and the presence of interactions.
The overlap between the nuclei with recent star-formation and the
presence of companions can be observed 
in Fig.\ref{fig.5}, where we present their distributions in the
histograms of Hubble types and  MGT types.

Many observations have provided evidence for a
causal link between strong nuclear starbursts and galaxy interactions,
which is also consistent with theoretical predictions.
The fraction of interactions in starburst galaxies
ranges from 20-30\% for the lower luminosity starbursts up to 70-95\%
for the higher luminosity ones (Kennicutt 1998 and references therein).
The frequency of companions in our whole sample is $\sim$30\%, but increases
to $\sim$60\% when we consider the subsample of composites,
which is close to that found among the most luminous starbursts.

The above result, combined with the evidence that the starbursts 
associated with interacting galaxies are also the youngest
(Cid Fernandes et al. 2001, GD01, SB00), supports an evolutionary
scenario for the relation between the nuclear starburst and
the AGN, as follows. First, galaxy interactions produce a 
flow of gas towards the center.
When the amount of gas piled up in the nuclear region is large enough, this 
gas, besides feeding the active nucleus, triggers a starburst.
The signatures of the interaction are still observable,  
the stellar population spectrum is of category 1,
and the emission-line ratios are intermediate between those of a 
Seyfert 2 and of a starburst (Cid Fernandes et al. 2001). 
The starburst then fades, and the
stellar population spectrum becomes dominated by older stars, being
observed as one of category 2 or 3, with a Seyfert 2 emission-line 
spectrum (Cid Fernandes et al. 2001). This evolution is also observed
in the MGT type, which changes from a late to an earlier type morphology.

The above scenario is also consistent with the results of numerical 
simulations of mergers (e.g. Henrquist \& Mihos 1995; Mihos \& Hernquist 1996). These simulations suggest that star-formation at the nucleus begins 
$\approx$500--800~Myrs after the beginning of interaction 
(galaxies closer than a few diameters), when clear signatures of the 
interaction are still visible. These signatures then
almost disappear in another 200~Myr or so, and only more subtle ones
remain, like small distortions and rings, such as those observed by Simkin, Su \& Schwarz (1980; see also Hunt \& Malkan 1999).

A similar evolutionary scenario has been recently proposed by Lei et
al. (2001) for LINERs. They have found that the
intensity of AGN activity in LINERs increases with decreasing 
star-formation contributions and suggest an evolutionary connection
from LINERs with strong star-formation and lower AGN activity
to those with no star-formation and stronger AGN activity.

\section{Summary and Concluding Remarks}

We have investigated the  relation between the nuclear stellar population 
properties -- in particular the incidence of recent star-formation -- and 
both the galaxy morphology and the presence of companions in a sample of 
35 Seyfert 2 galaxies. The results found for the Seyferts were compared
with those of two control samples of non-Seyfert galaxies.

The main conclusions of this work are:

The Hubble types of the Seyfert 2 galaxies are evenly distributed
from S0 to Sb, then the number of Sc galaxies drops to less than half 
the number of galaxies within each of the earlier Hubble types.
When compared with a control sample of non-Seyfert galaxies,
the present Seyfert 2 sample shows a $\sim$50\% deficiency of Sc galaxies,
and an excess of galaxies with uncertain or peculiar morphology (although
this latter result is apparently due in part to the difficulty in ascertaining
a Hubble type to the most distant galaxies of the sample).
This conclusion reinforces the known result that Seyfert nuclei are preferentially found in earlier type hosts (Ho, Filippenko \& Sargent 1997).

The number of Seyfert 2 galaxies 
with composite nuclei (Seyfert + Starburst) increases
towards the later Hubble types. The fraction of 
galaxies with recent star formation
is similar to that found in non-Seyfert galaxies for the Hubble
types Sb or later, but seems to be larger in Seyfert 2 nuclei
for the earlier Hubble types. 

The nature of the ambiguous blue continuum PL/YS 
is a key issue in assessing the extent of
the difference between the stellar population of Seyfert 2 nuclei
and normal galaxies of the same Hubble type. If this continuum is 
at least in half the cases due to young stars, then the fraction of
Seyfert 2 galaxies with recent circumnuclear star formation 
would be larger than that in normal galaxies for all Hubble types.
This ambiguity can only be solved with high signal-to-noise ratio 
UV or near-UV spectra obtained at high spatial resolution,
observations which are presently feasible with the Hubble Space Telescope.  

The number of Seyfert 2 nuclei dominated by an old stellar population increases
towards the early Hubble types, similar to what is found in normal 
galaxies. Nevertheless, within each of the Hubble types S0, Sa and Sb, the
fraction of Seyfert 2 nuclei dominated by the old stellar population is 
systematically smaller than that in the normal galaxies, due to the 
larger fraction of Seyfert 2 galaxies with recent star-formation and  PL/YS
continuum in these Hubble types.

There is a very good correlation between the presence of recent star formation
and a late ``inner Hubble type'', assigned by MGT to the
galaxies of our sample based on high spatial resolution
HST F606W images. Our interpretation for
this correlation is a larger gas content in the Seyfert 2 
galaxies with recent star formation in and around the nucleus, which,
through the associated dust, makes the 
gas distribution noticeable in the F606W images.
This conclusion is consistent with the results of Cid Fernandes et al. 
(2001), who have found a correlation 
between the young stellar content and the infrared 
luminosity of the galaxy in the same sample, supportint a larger amount of
dust emission in the Seyfert 2 galaxies with recent star formation. 

Another good correlation is found between the presence of companions and the
incidence of recent star formation in the Seyfert 2 nuclei.
The frequency of companions in our whole sample is 30\%, but increases
to 60\% when we consider the subsample with recent  star-formation.
Combined with the fact that the interacting galaxies are the
ones with the youngest stellar population, this result suggest
an evolutionary scenario in which the interaction is responsible
for sending gas inwards, which both feeds the AGN and triggers 
star-formation, giving origin to a composite nucleus. The 
Starburst then fades with time and the composite nucleus turns into
a ``pure'' Seyfert 2 nucleus with an older stellar population.
This scenario can reconcile the hypothesis that interactions
are responsible for triggering nuclear activity in Seyfert
galaxies with previous observational studies which do not find a large
excess of interacting galaxies in Seyfert
samples when compared with non-Seyfert ones. Signatures of the interactions
are only clearly observed in the initial stages, which coincides
with the phase in which a composite (Seyfert+ Starburst) nucleus
is observed.

\acknowledgements
We are pleased to thank the hospitality of the INAOE, Tonantzintla, Mexico,
and in particular Itziar Aretxaga and Daniel Kunth, 
during the Guilhermo Haro workshop of July, 2000,
when this work was initiated. We also thank the
suggestions by the referee which helped to improve the paper.
We acknowledge support from the
brazilian institutions CNPq, CAPES and FAPERGS.
HRS work was partially supported by 
NASA under grant No. NAG5-9343.
We have made use of the NASA/IPAC
Extragalactic Database, operated by the Jet Propulsion Lab, Caltech, 
under contract with NASA. The National Radio Astronomy Observatory 
is a facility of the National Science Foundation operated under 
cooperative agreement by Associated Universities, Inc..


\begin{deluxetable}{lcccccc}
\tabletypesize{\scriptsize}
\tablecaption{Sample properties\tablenotemark{a}}
\tablewidth{0pc}
\tablehead{
&&&\multicolumn{1}{c}{scale}&\multicolumn{2}{c}{Hubble type}&\\
\cline{4-4}\cline{5-6}
\colhead{Galaxy}&
\colhead{$V_{GSR}$}&
\colhead{$M_{B_T^0}$}&
\colhead{pc/arcsec}&
\colhead{RC3}&
\colhead{MGT}&
\colhead{Pop.}}
\startdata
Southern sample &&&&&&\cr
NGC\,1358        &3980      &-20.92       &257.3	&SB0/a &SB0    &2 \cr
NGC\,1386        &741       &-19.02       &81.9\tablenotemark{1}	&SB0   &Sb/c   &2 \cr
NGC\,3081        &2164      &-19.71       &139.9	&SB0/a &SB0/a  &3 \cr
NGC\,5135        &3959      &-21.24       &255.9	&SBab  &Sc     &1 \cr
NGC\,5643        &1066      &-20.53       &68.91	&SABc  &-      &1 \cr
NGC\,6300        &997       &-20.42       &64.45	&SBb   &Sc     &2 \cr
NGC\,6890        &2459      &-19.76       &158.9	&Sb    &-      &3 \cr   
NGC\,7130        &4850      &-21.17       &313.5	&Sa    &Sd     &1 \cr
NGC\,7582        &1551      &-20.75       &100.3	&SBab  &?      &1 \cr
Mrk\,348         &4669      &-20.03       &301.8	&S0/a  &S0     &3 \cr
Mrk\,573         &5161      &-20.62       &333.6	&SB0   &S0     &3 \cr
MRK\,607         &2716      &-19.48       &175.6	&Sa    &Sb     &2 \cr
MRK\,1210        &3910      &-19.38       &252.7	&Sa\tablenotemark{2} &Sa   &1 \cr
CGCG\,420-015    &8811      &-20.35       &569.5	&Sa\tablenotemark{2} &Sa   &2 \cr
IC\,1816         &5086       &-20.5       &328.8	&Sab   &SBa/b   &3 \cr
IRAS\,11215-2806 &4047      &-20.66       &261.6	&S0\tablenotemark{2} &S0  &2\cr
MCG-5-27-13      &7263      &-21.22       &469.5	&SBa   &Sb     &3 \cr
Fairall\,316     &4772      &-20.21       &308.5	&S0    &S0     &2 \cr    
ESO\,417-G6      &4792      &-19.88       &309.8	&RS0   &-      &2 \cr
ESO\,362-G8      &4616      &-20.44       &298.4	&S0    &Sa     &1 \cr
Northern sample\tablenotemark{3} &&&&&&\cr
Mrk\,1           &4970      &-19.66       &321.3	&S     &Sc     &1 \cr
Mrk\,3           &4124      &-20.31       &266.6	&S0    &S0     &3 \cr
Mrk\,34         &15140\tablenotemark{2} &-21.36\tablenotemark{4}  &978.6	&S\tablenotemark{2} &-  &3 \cr
Mrk\,78         &11288      &-20.87\tablenotemark{4}       &729.7	&SB\tablenotemark{2} &- &1 \cr
Mrk\,273        &11390      &-21.11\tablenotemark{4}       &736.2	&Ring galaxy\tablenotemark{2} &- &1 \cr 
Mrk\,463E       &15209       &-22.5       &983.1	&?     &-     &1 \cr
Mrk\,477        &11511       &-20.7       &744.1	&Sp    &-     &1 \cr 
Mrk\,533         &8912      &-21.76       &576.1	&Sbc   &S(B)c &1 \cr
Mrk\,1066        &3705      &-20.43       &239.5	&SB0   &Sc    &1  \cr
Mrk\,1073        &7097      &-21.98       &458.8	&SBb   &Sc    &1  \cr
NGC\,1068        &1144      &-21.45       &73.95	&Sb    &-     &3 \cr
NGC\,2110        &2153      &-20.52\tablenotemark{4}       &139.2	&SB0   &Sa    &2 \cr
NGC\,5929        &2684      &-20.17\tablenotemark{4}       &173.5	&Sab   &S0    &2 \cr 
NGC\,7212        &7972      &-21.12       &515.3	&S?    &Irr?  &3 \cr
IC\,3639         &3137       &-20.4       &202.8	&SBbc  &SBb   &1 \cr 
\enddata
\tablenotetext{a}{Columns: 
(1) Velocity in the Galactic Standard of Rest (from RC3);
(2) Absolute blue magnitude using B$_{T^0}$ from RC3;
(3) Scale in parsecs per arcsec;
(4) Hubble type as in RC3;
(5) Inner Hubble type from MGT;
(6) Stellar population category: 1) young stellar population, 2) old stellar population, 3) blue light of uncertain origin.}
\tablenotetext{1}{Adopting distance to the Fornax cluster of 16.9\,Mpc (Tully 1988)}
\tablenotetext{2}{From NED}
\tablenotetext{3}{Excluding galaxies in common with the southern sample}
\tablenotetext{4}{Using $B_T^0$ from Whittle (1992)}
\end{deluxetable}

\begin{deluxetable}{lcccccccc}
\tabletypesize{\scriptsize}
\tablecaption{Subsample of interacting Seyferts\tablenotemark{a}}
\tablewidth{0pc}
\tablehead{
\colhead{Seyfert}&
\colhead{$V_R$}&
\colhead{$M_{B_T^0}$}&
\colhead{Companion}&
\colhead{Distance}&
\colhead{$V_R$}&
\colhead{$\Delta M_B$}&
\colhead{Pop.}&
\colhead{Ref.}}
\startdata
NGC\,5135  &4112 &-21.24 &IC\,4248\tablenotemark{2}   &212\,SE   &4133  &0.97  &1 &Kollatschny \& Fricke 1989 \cr                    
NGC\,7130  &4842 &-21.17 &Unidentified\tablenotemark{3} &15.7\,NW   &- &-  &1 &Gonz\'alez Delgado et al. 1998\cr   
NGC\,7582  &1575 &-20.75 &NGC\,7590  &59.1\,NE &1596   &0.74  &1 &Kollatschny \& Fricke 1989\cr
Mrk\,348   &4507 &-20.03 &NPM1G+31.0016 &21.7\,E   &-  &1.89 &3 &Rafanelli et al. 1995\cr
MRK\,607   &2716 &-19.48  &NGC\,1321   &16.9\,N    &2698   &0.89 &2 &Colbert et al. 1996\cr                     
Mrk\,1     &4780 &-19.66  &NGC\,451  &36.6\,SE &4880 &-0.12  &1 &Rafanelli et al. 1995\cr
Mrk\,273   &11326 &-21.11 &Merger: double nucleus &- &- &- &1 &Mazzarella \& Boroson 1993\cr
Mrk\,463E  &14990 &-22.5  &Merger: double nucleus &- &- &- &1 &Heisler \& Vader 1994 \cr
Mrk\,477   &11332 &-20.7  &KUG1439+537  &37.5\,NE  &-  &- &1 &De Robertis 1987\cr
Mrk\,533   &8713  &-21.76 &NGC7674A\tablenotemark{2}  &20.7\,NE  &8852  &2.18 &1 &Verdes-Montenegro et al. 1997\cr
NGC\,5929  &2492  &-20.17 &NGC\,5930  &5.2\,NE &2672  &-0.50 &2 &Gonz\'alez Delgado et al. 1997 \cr
NGC\,7212  &7984  &-21.12 &NGC\,7172\,NED03\tablenotemark{2} &9.3\,NE  &8167  &-  &3 &Veilleux et al. 1997\cr 
IC\,3639   &3275  &-20.4  &ESO\,381-G09\tablenotemark{2}  &21.9\,NE  &3050 &0.89 &1 &Gonz\'alez Delgado et al. 1998\cr   
\enddata
\tablenotetext{a}{Columns:
(1) Name of the Seyfert;
(2) Heliocentric velocity from NED;
(3) $M_{B_T^0}$ as in Table 1;
(4) Name of the companion;
(5) Distance between Seyfert and companion in kpc using angular separation
from NED;
(6) Heliocentric velocity of the companion from NED;
(7) Difference: magnitude of the companion minus that of the Seyfert from NED;
(8) Stellar population category of the Seyfert; 
(9) Previous reference on the companion(s).}
\tablenotetext{2}{belongs to a group; data is for the closest companion}
\tablenotetext{3}{In the DSS there seems to be two dwarf galaxies close to IC5135;
images by Shields and Filippenko (1990) show tidal distortions
and confirm the presence of the small unindentified galaxy to NW.}
\end{deluxetable}

\end{document}